\documentclass[aps,prl,twocolumn]{revtex4}
\usepackage{color}
\usepackage{graphicx}
\usepackage{dcolumn}
\begin{document}

\title{On the giant enhancement of light in plasmonic or all-dielectric gratings comprising nano-voids}

\author{J\'er\^ome Le Perchec}\email{jerome.le-perchec@cea.fr}

\affiliation{CEA, LETI, Minatec Campus, Optics and Photonics Department, 17 avenue des Martyrs, 38054 Grenoble, France}

\date{\today}
\begin{abstract} 
We report the possibility to generate tremendous light-field enhancements within shallow nano-trenches made in a high index dielectric material, because of resonant behaviours reminiscent of what we get with sub-wavelength plasmonic cavities. The high quality factors are explained through a modal analysis and can be tuned with appropriate design rules. The thin dielectric void gratings here simulated could be a relevant alternative to plasmon-based devices for chemical sensing, or could be used as efficient wavelength-selective photo-absorbers by taking weakly absorbing materials.
\end{abstract}


\maketitle
Managing light at sub-wavelength scales by means of guiding, confinement, and amplification, is of central interest for numerous concrete applications (bio-sensors, advanced photo-detectors, telecommunication, micro displays) and should remain the object of an intense research activity in the future. To overcome the diffraction limit, coupling of light with electromagnetic (EM) modes showing a high effective index may offer new and different propagation properties. Plasmonics is one of the well-known branch which explores this coupling with nanostructured metallic surfaces, especially aiming at light concentration in small volumes \cite{Gramotnev,Schuller,Kazemi}. Moreover, the latter surfaces can play the concomitant role of electrodes if needed. In parallel, all-dielectric (all–photonic) structures based on high optical index materials have become an ever-increasing field of study since the last decade \cite{Almeida,Karagodsky,Jahani} showing very interesting properties as well. They have to be distinguished from classical photonic crystals. 
An important point is they are free of the dissipation losses that are inherent to surface plasmon-polariton (SPP) waves, provided we meet some technological fabrication challenges (the sidewall roughness of dielectric waveguides should be minimized for instance \cite{Kita}). Also, by mixing plasmon modes and high index materials, some trade-off is enabled to design more robust or efficient optical devices like hybrid waveguides \cite{Oulton,Alam} or resonators \cite{Yang}, compact photo-absorbers \cite{leperchec09,Cui} (oftenly inspired from antennas), or versatile transmission optical filters \cite{LePerchec11} compatible with large scale microelectronics silicon processes.
Actually (and it can be understood from a mathematical point of view), some resonant behaviours obtained for structures made in a negative permittivity material (metals, or ionic crystal in the reststrahlen band) can also similarly hold by taking a highly positive dielectric permittivity, although EM modes and excitation mechanisms are not the same\cite{Jahani,Li,Devilez}. Consequently, depending on wanted specifications (spectral bandwidth, incidence angle sensitivity), high-index-contrast architectures may constitute solutions possibly better than plasmonic ones. In this paper, we exemplify this statement by numerically and analytically showing the light concentration within strongly sub-wavelength and \textit{weakly deep} rectangular trenches made in a dielectric material, and highly resembling the metallic nano-resonators reported in \cite{LePerchec08,Polyakov}. Controlling hot spots on predefined architectures is of special importance for chemical sensing (fluorescence enhancement or Raman spectroscopy), all the more that transparent substrates would avoid parasitic re-absorption of the useful emitted signal \cite{Liu10,Kita}. Besides, in case of a weakly absorbing semi-conductor, sharp resonances may persist and yield strong optical absorption. 

\begin{figure}[h]
\begin{center}
\includegraphics[angle=0,scale=.40]{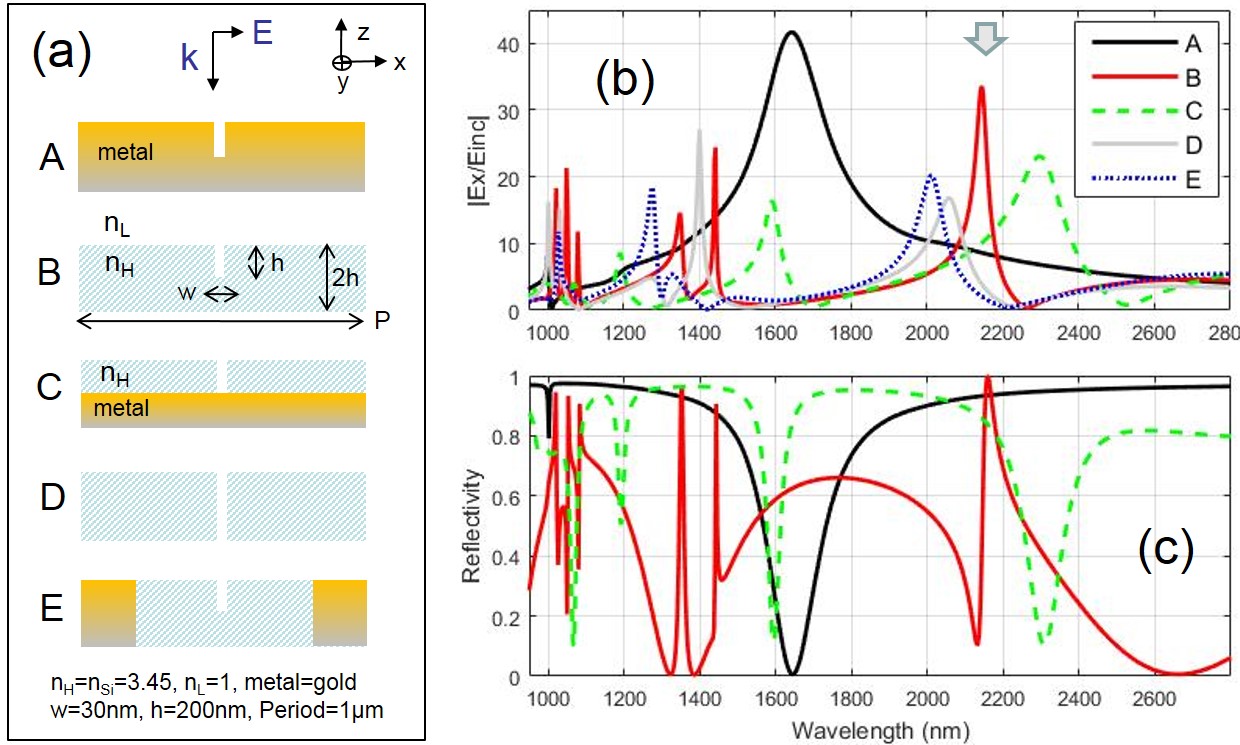}
\end{center}
\caption{(color online)  (a) Sketches of different generic void gratings presenting a high permittivity contrast, either metallic (gold), either all-dielectric (index $n_H =\sqrt{\varepsilon_H}= n_{Silicon}$), or hybrid. External medium is air ($n_L=1$). H means high, and L low. Geometrical parameters are fixed. In case (E), the metallic part is 200nm width. (b) Normalized electric field modulus inside the voids, in function of $\lambda$, at normal incidence. (c) Reflectivity spectra for the first three structures.}\label{figure1}
\end{figure}

\begin{figure}[h]
\begin{center}
\includegraphics[angle=0,scale=.41]{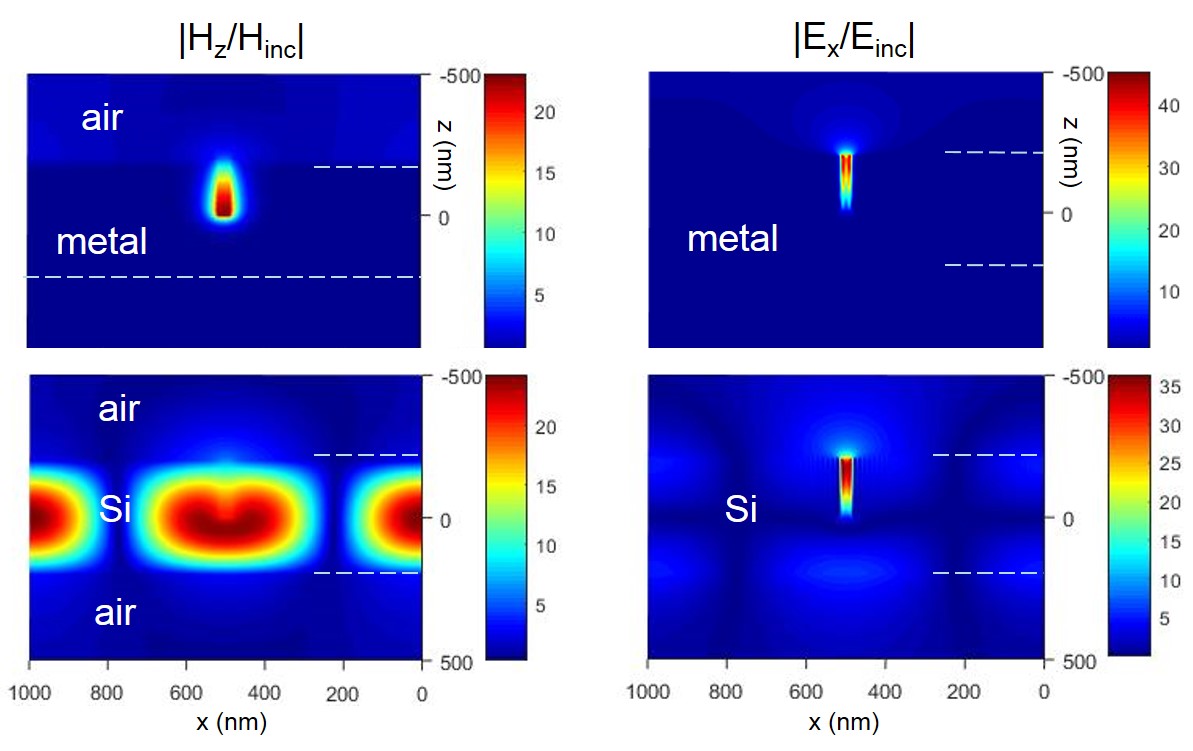}
\end{center}
\caption{(color online) Maps of the magnetic and electric field modulus at the resonance wavelengths of the structures (A) ($\lambda=1.65\mu m$) and (B) ($\lambda =2.145 \mu m$) described in Fig.~\ref{figure1}(a).}\label{figure2}
\end{figure}

As a first approach, let us draw a comparative study of strongly sub-wavelength one-dimensional (1D) cavity gratings showing high permittivity contrasts, whose constitutive materials may be gold ($\Re(\varepsilon)<<0$), silicon ($\varepsilon=11.9$) or a mixture of them, and for a fixed set of geometrical parameters, in the short infrared range (see Fig.~\ref{figure1}(a)).  The excitation wave is transverse-magnetic (TM) polarized and the wavelength is $\lambda$. Special attention will be paid regarding the electric field amplitude inside the voids. Simulations are carried out through Rigorous Coupled Wave Analysis (RCWA) taking a periodicity $P=1\mu m$.
Figure~\ref{figure1}(b) shows spectra of the normalized electric field modulus calculated near the mouth of the cavities, for the five different configurations of Fig.~\ref{figure1}(a).  In parallel, Figure~\ref{figure1}(c) gives the corresponding reflectivity responses for the first three cases (the other ones are not displayed for readability reasons and do not  bring substantial information). We focus us on the peaks observed around $2.1\mu m$ (fundamental resonances), for which $w\sim\lambda/70$ and $h\sim\lambda/10$. The full-metal structure (A) exhibits a well-known Fabry-Pérot (FP) resonance (which is also a Fano resonance) supported by the cavity plasmon mode. This is correlated with a total optical absorption inside the nano-voids.  The full-silicon groove structure (B) also shows a quantitatively close enhancement peak (intensity $|E_x/E_{inc}|^2 >10^3$ at $\lambda=2.145 \mu m$), and a better quality (Q) factor. The far-field response is very different (no absorption), with a total reflection effect very close to this resonance wavelength. Maps of Fig.~\ref{figure2} clearly shows that we  get quite analogous electric field patterns in cases (A) and (B), indicating that (B) also supports a type of vertical FP resonance (we will give details on it later). The hybrid structure (C) is inspired from the electric field symmetry observed on (B) (the horizontal plane $y=0$ is nodal) and still supports a FP resonance which is a little bit damped and spectrally broadened because of the metal substrate.
Structure (D) forms a high index bar grating (somewhat similar to that seen in \cite{Karagodsky} but much more thinner here), and the electric field pattern becomes symmetrical on both part of the plane $y=0$ at the resonance (not shown). We note in Fig.~\ref{figure1}(a) that for higher order resonances ($\lambda \approx 1.4\mu m$), the field peak of (D) can overpass the (B) one and the Q factors may be improved too. At end, (E) derives from (B) where a metallic wall 200nm width is inserted in the high index region at mid-distance of the cavities. Here again, an enhancement peak persists, showing that the resonance does not need a continuous propagation of possible diffractively guided modes all along the dielectric plate. However, one can find that, by enlarging the metallic part (at fixed period), the resonance disappears if the dielectric region width becomes lower than $\lambda/n_H$.

Let us get more insight in configurations (A) and (B) to clarify the similarities and differences with respect to the so-called vertical FP resonances. In the framework of a modal approach, at \textit{normal} incidence, the fundamental mode in the grating (and every even eigen function with respect to the x variable) has a wave vector $k_z=k n_{eff}$ given by the following transcendental equation:
\begin{equation}\label{eqtranscendental}
k_x^L \varepsilon_H \tan \left( \frac{k_x^Lw}{2} \right) + k_x^H\varepsilon_L \tan \left( \frac{k_x^H (P-w)}{2} \right) = 0
\end{equation}
with $k_x^{L,H}=k(\varepsilon_{L,H}-n_{eff}^2)^{1/2}$ and $k=2 \pi/\lambda$. Taking a good metal ($\varepsilon_H \ll 0$), we find that the cavity mode corresponds to a guided plasmon wave characterized by:
\begin{equation}\label{neffmetal}
n_{eff}^2 \approx \varepsilon_L \left( 1+\frac{2\delta}{w} \tanh \left(\frac{P-w}{2\delta} \right) \right)
\end{equation}
where $\delta$ is the classical metal skin depth.  In case (A) of Figure 2, $n_{eff}=1.64+i0.031$ (numerical value given by RCWA, well consistent with Eq.(2)). For strongly sub-wavelength cavities,  $n_{eff}$ dramatically increases (with no cut-off), and provided the period is not too small, the waveguide mode may enter the electro-static regime \cite{LePerchec08} characterized by a strong electric component and a weak magnetic field ($E_x/H_z  \propto n_{eff}$). As a consequence, optical hot spots (local intensity enhancements of several thousands and much more) based on FP resonances may be generated inside relatively shallow nanometric grooves or notches (following a relation $\lambda_{res} \propto 4h n_{eff}$). The intensity enhancements are all the more important as the geometrical concentration ratio $(P/w)^2$ is high (the Q factor may be optimized with h).

\begin{figure}[h]
\begin{center}
\includegraphics[angle=0,scale=.41]{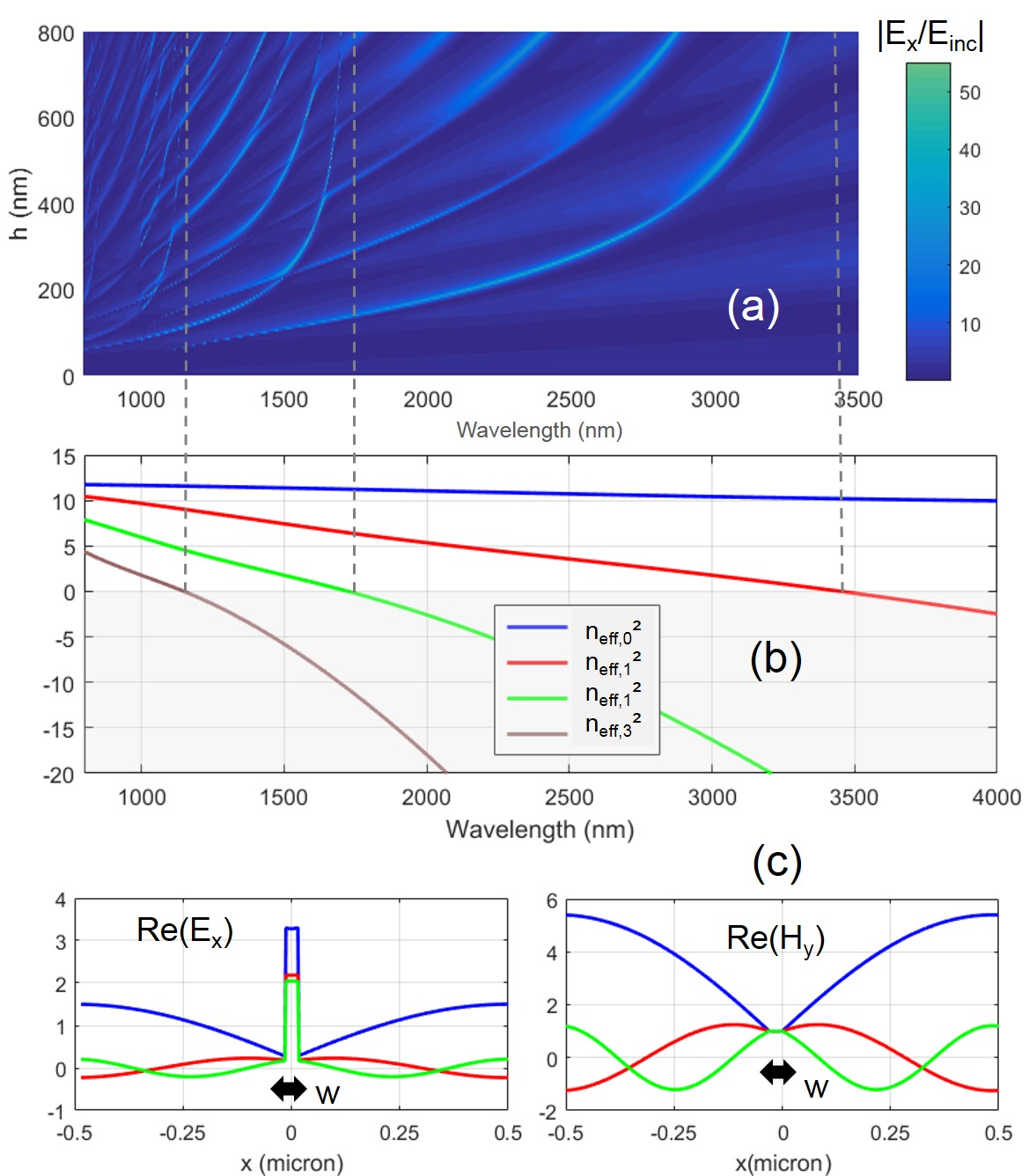}
\end{center}
\caption{(color online) (a) Diagram of the field enhancement inside the void of structure (B) in function of $\lambda$ and depth h. (b) Evolution of the square effective indexes of the four first eigen modes respecting Eq.(\ref{eqtranscendental}) depending on $\lambda$ (when they are negative, the modes are said evanescent along z-direction). (c) Exact field profiles of the first three eigen modes (even functions) of the dielectric grating at $\lambda=2.145 \mu m$ (resonance seen in Fig.~\ref{figure2}). The void is centered at $x=0$.}\label{figure3}
\end{figure}

Now, consider the entirely dielectric grating, with $\varepsilon_H>10$ and $\varepsilon_L\sim 1$ typically. A general remark is that, according to boundary conditions in TM polarization, $E_x^L  / E_x^H= \varepsilon_H/\varepsilon_L>1$ at the vertical walls, so that a cavity mode is expected to have a boosted electrical component \cite{Robinson} in a narrow void (as for the metal case). Moreover, we know that $n_{eff}^2<max(\varepsilon_H,\varepsilon_L)$ whatever the dielectric structure. \textit{What exactly supports the FP modes and how explain the strength of these resonances} (in order to improve the design)? The lossless property of all-dielectric structures is not a sufficient resaon for high Q factors because radiation leakage is another important limiting mechanism. We will begin with the first part of the here-above question. Taking again the eigenvalue equation (\ref{eqtranscendental}) with $w \ll P<\lambda$, the effective index of the fundamental mode obeys the relation:
\begin{equation}\label{neff0}
n_{eff,0} \approx \frac{n_H}{\sqrt{1+\frac{w}{P}(\frac{\varepsilon_H}{\varepsilon_L} -1)}} 	 	
\end{equation}
It weakly depends on $\lambda$ (contrary to the metal case) and has no cut-off (as for the metal). In case (B) Fig.~\ref{figure2}, $n_{eff,0}= 3.31$. For a greater index contrast $n_H/n_L$, a narrower w or a greater P is needed to maximize the local effective index. Looking at the electric field pattern in Fig.~\ref{figure2}, it could seem this slot mode is responsible for the strong FP resonance. But it is actually \textit{not} the case: this is the second eigen mode, whose field profile is also close to that of the fundamental one, as we will see. Indeed, we said that if the dielectric region width was smaller than $\lambda/n_H$, no strong resonance occurs. Once $\lambda<n_H (P-w)$, a second eigen mode develops a positive effective index which almost corresponds to the lowest-spatial-frequency Bragg condition $\tan(k_x^H (P-w)/2)\approx 0$ (see Eq.(\ref{eqtranscendental})), i.e.:			
\begin{equation}\label{neff1}
n_{eff,1} \alt n_H\sqrt{1-\left(\frac{m\lambda}{n_H (P-w)}\right)^2}
\end{equation}
with $m=1$ here, and $ n_{eff,1}< n_{eff,0}$. In case (B) of Fig.~\ref{figure2}, $ n_{eff,1}= 2.19$ (exact numerical solution).  To confirm and better understand what happens, pictures in Fig.~\ref{figure3} are very illuminating. The diagram of the electric field amplification inside the void has been calculated in function of wavelength and cavity depth (P and w are still fixed). In parallel, Fig.~\ref{figure3}(b) and (c) give the effective indexes and profiles of the grating eigen modes. When $2n_HP< \lambda<n_HP$, we find a first set of harmonic vertical FP resonances following a law $h \propto p\lambda/4 n_{eff,1}$ where p is a positive integer. The field maps in Fig.~\ref{figure2} (silicon case) are clearly consistent with the second-order mode profile in Fig.~\ref{figure3}(c). When $3n_HP<\lambda <2n_HP$, a third eigen mode becomes propagative ($n_{eff,2}>0$), and an additional set of harmonic FP resonances appears. These sets may interfere (that is why we see a multiplication of resonances in Fig.~\ref{figure1}(a) and (b) for shorter $\lambda$). When $\lambda<P$, a new diffraction order becomes propagative in the external (air) medium and all the enhancements suddenly drop because of radiation leaks. At last, the fundamental slot mode is only associated to bad quality resonances identified as classical half-wave plate conditions.  It is interesting to observe that we can get resonances for weakly deep cavities (up to 70nm depth only in this example), while the Q factor (inverse of spectral width) is generally better for thick gratings. 
At this stage, the question of the mechanism behind the high-Q factors is still to be answered. In Ref.\cite{Karagodsky}, the authors reported RCWA simulations of high-contrast and thick bar gratings, in TE polarization, and interpreted the high-Q factor resonances as the result of strongly coupled simultaneous FP resonances of (at least) two waveguide-array modes. In the present paper, we give a different and maybe more explicit explanation. According to the exact modal method \cite{Botten, LePerchec08} (different from RCWA where modes are expanded onto a Fourier basis), true eigen modes form an orthogonal basis so there is no direct coupling between them. If one considers the case of a reflecting grating at normal incidence (as the structure (C) in Fig.~\ref{figure1}(a), but taking a perfect mirror), algebraic calculations are simplified and one shows that each mode amplitude behaves as:
\begin{equation}\label{Am}
A_m^{-1}\propto \cot(kh n_{eff,m}) -\frac{i n_{eff,m}}{<F_m|F_m>} \sum_{q=-\infty}^{+\infty} \frac{ |<e_q|F_m>|^2}{\sqrt{1-(q\lambda/P)^2}}
\end{equation}
where the coupling terms are defined by:
\begin{equation} \label{coupling}
<f|g>=\frac{1}{P} \int_{0}^{P} f^* (x) g(x) \frac{1}{\varepsilon(x)} dx 	
\end{equation} 
$F_m$ is the $m^{th}$ eigen function describing the \textit{magnetic} field component $H_y$ in function of x (see Fig.~\ref{figure3}(c)). $(e_q)_q=(e^{i 2\pi q x/P})_q$ is the usual P-periodic Rayleigh basis describing the field outside the grating. The FP resonance condition occurs when the real part of the denominator of $A_m$ nearly cancels. As $<F_m|F_m>$ is necessarily positive for dielectrics, and $P<\lambda$, the remaining part in the denominator reduces to:
\begin{equation}\label{rdet}
r=-i~n_{eff,m}\frac{ |<1|F_m>|^2 }{<F_m|F_m>} 	
\end{equation}
We have thus a direct vision of what bounds the FP resonance: it essentially depends on the magnetic field profile (we are in TM-polarization). Given $w/$P is very small \textit{and} eigen modes with $m>0$ show a more oscillating behavior inside the high index dielectric parts (see Fig.~\ref{figure3}(c)), their contribution to the limiting term r is much lower than that of the fundamental slot mode. It means that their coupling to the zero order diffraction ray, or their radiation leak outside the grating, is weak. In other words, high-contrast index horizontal interfaces play the role of good mirrors for these internal waves, hence the resonance. To sum-up, at view of Eq.(\ref{rdet}), a first way to get efficient FP resonances is to work with a spatially oscillating propagative eigen mode, excited through a small resonator aperture (w), by taking a period (i.e. a dielectric region size) sufficiently \textit{great} in the spectral range of interest. Another way is to resort to small $n_{eff}$  (near-cutoff mode) not far from a Bragg condition in our periodic case, but resonances are only obtained for very deep cavities (as seen in Fig.~\ref{figure3}(a)), which is technologically difficult to make (the Deep Trench Isolation technique known in silicon microelectronics could be used \cite{Trivedi,Tournier}). By comparison, for a perfect metal ($n_{eff}=1$, electromagnetic field null in the metal), Eq.(\ref{rdet}) reduces to $r=-iw/P$. For a real metal, $n_{eff}>1$ and develops an imaginary part introducing an additional (absorption) term. Strong resonances may not hold for a 2D structure with voids having a small rectangular cross-section in the $(x,y)$-plane, because the modes would tend to be evanescent (excepting the fundamental one), and the coupling with the incident wave would become inefficient. Instead, crossed 1D (infinite) trenches work well and allow independence with respect to light polarization.

\begin{figure}[h]
\begin{center}
\includegraphics[angle=0,scale=.47]{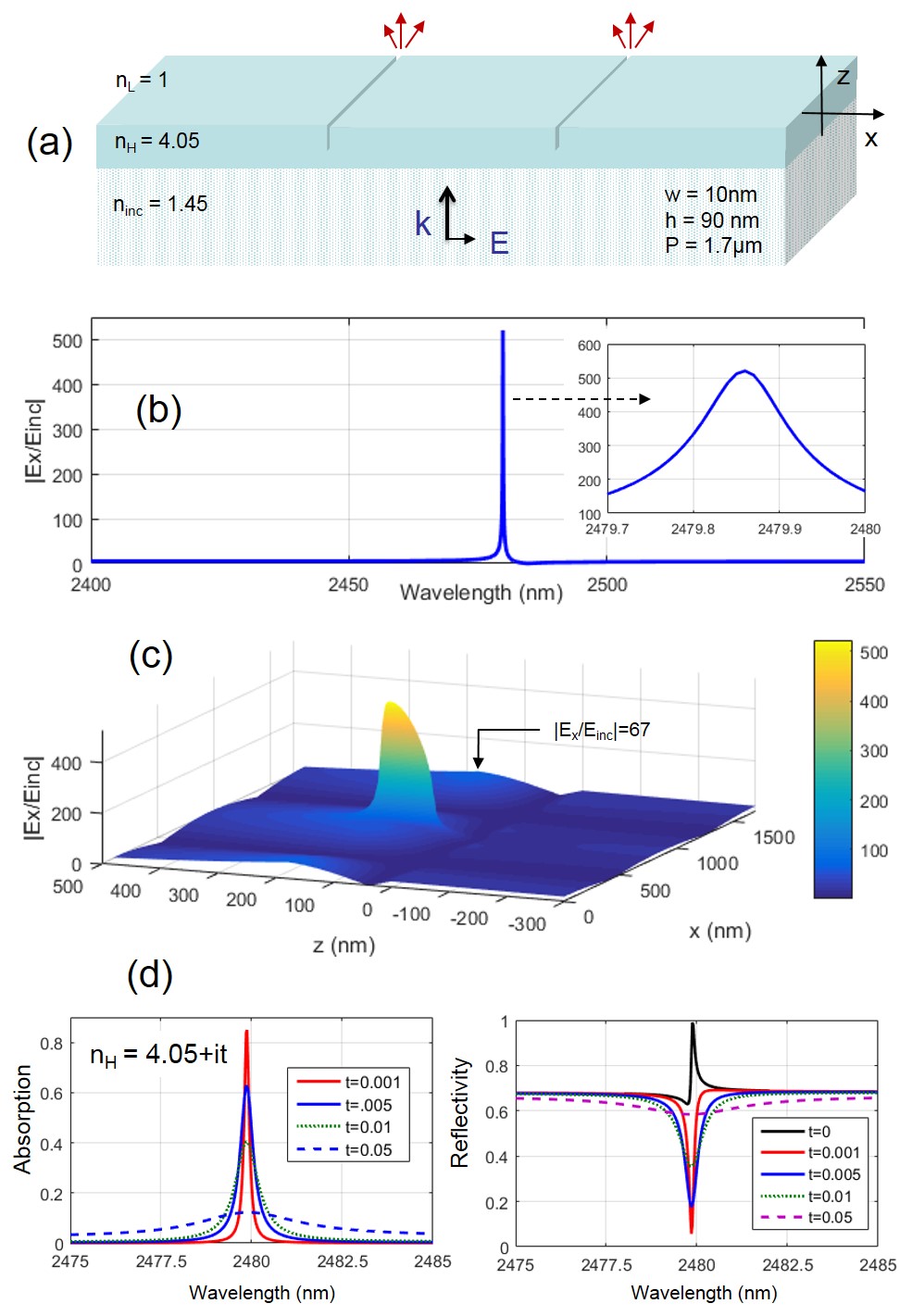}
\end{center}
\caption{(color online) (a) Optimized Germanium structure with voids 90nm deep and excited on its backside, in the mid infrared. (b) Corresponding field spectrum inside the voids, showing a sharp FP resonance. (c) Fully computed electric field map at $\lambda=2479.86 nm$ over one period ($n_{eff,1}=3.58$, $\lambda/h=27.55$). (d) Photo-absorption and reflectivity spectra when an imaginary part is added to the high index.}\label{figure4}
\end{figure}

Taking advantage of our previous analysis, one can generate giant field enhancements in tiny (not too deep) indentations. We propose the structure designed in  Fig.~\ref{figure4}(a), based on a higher permittivity material (Germanium, with $n_H \approx 4.05$ for $\lambda>2\mu m$). The thin grating is supported by a bulk silicon oxide substrate, and is excited on the backside, which differs from the classically transverse excitation of grating waveguides for chemical or biological sensing \cite{Kita}. Here, the structure could be used both in reflection or transmission configurations, where molecules would be deposited on the free surface, especially into the voids to amplify the signal.  Figure~\ref{figure4}(b) clearly shows a very sharp photonic resonance. It corresponds to a huge electric field intensity $|E_x^L/E_{inc} |^2>2.7\times 10^5$ close to $\lambda=2.48\mu m$ (convergent values given by RCWA with 500 Fourier orders), inside voids whose depth $h\approx\lambda/27$ only. The corresponding Q factor is $1.65\times 10^4$. The intensity enhancement just above the horizontal free surface is also high ($\sim 4.10^3$, see Fig.~\ref{figure4}(c)), which strongly differs from a metallic case where the field is only amplified inside cavities. This evancescent surface field is quite delocalized onto the whole period and stocks reactive power \cite{Leperchec10}.

Finally, by introducing a small imaginary part ($it$) to $n_H$, as shown in Fig.~\ref{figure4}(d), we observe that a high-quality resonance may be preserved. It leads to strong frequency-selective photo-absorptions (up to 85\%), without shifting the excitation wavelength. The peak only vanishes when $t=0.05$,  showing some robustness of the resonance. The far-field response also exhibits a clearly detectable (absorption) signature. A thin layer of indirect band-gap semiconductor could thus be rendered anomalously absorbing by making a series of narrow indentations.  It is worth pointing out that, for a fixed geometry a fine \textit{tuning} of the resonance frequency may be done by tilting the incidence angle of a fraction of degree (or more) as it slightly modifies the effective index (note that oblique incidence may also excite additional odd eigen modes at other wavelengths). Besides, if this angle is above the critical one in the input medium (or if the system is excited by an evanescent wave as for metallic gratings \cite{Leperchec10,Quemerais}), we could kill or minimize radiation leaks and get a gain in Q factor (not shown here). One can imagine many other configurations (multiple cavity arrangements, for instance). V-shaped dielectric grooves should also work, and higher index materials, like PbSe or Te in the infrared, could lead to extreme phenomena. 

In conclusion, it has been theoretically shown that very sub-wavelength rectangular voids made in a high index dielectric medium can support giant electric field intensities, thanks to strong photonic Fabry-Perot resonances that remind those obtained with plasmonic nano-grooves, and even for depths sometimes several tens times lower than $\lambda$. The oscillating eigen modes supporting such resonances and the reason for very high Q factors have been explained through a modal analysis. Design rules have thus been indicated. The point is to work with the right discrete mode, provided appropriate boundary conditions exist. As an alternative to SPP-based sensors, such high-enhancement-factor transparent surfaces could be well used for molecule excitation and detection. Taking a slightly absorbing semiconductor, efficient resonances may persist and be used for wavelength-selective photo-absorbers based on thin active layers. These results should motivate experimental works.


\begin{thebibliography}{99}
\bibitem{Gramotnev} D.K. Gramotnev and S.I. Bozhevolnyi, Nat. Photonics \textbf{4}, 83 (2010).
\bibitem{Schuller} J.A. Schuller \textit{et al}, Nat. Materials \textbf{9}, 193–204 (2010).
\bibitem{Kazemi} N. Kazemi-Zanjani, S. Vedraine, and F. Lagugn\'e-Labarthet, Opt. Express \textbf{21}, 25271 (2013).
\bibitem{Almeida} V. R. Almeida \textit{et al}, Opt. Lett.\textbf{29}, 1209 (2004). 
\bibitem{Karagodsky} V. Karagodsky and C.J. Chang-Hasnain, Opt. Express \textbf{20}, 10888 (2012).
\bibitem{Jahani} S. Jahani and Z. Jacob, Nat. Nanotechnol. 11, \textbf{23} (2016).
\bibitem{Kita} D.M. Kita, J. Michon, S.G. Johnson, and J. Hu, Optica \textbf{5}, 1046 (2018).
\bibitem{Oulton} R.F. Oulton \textit{et al}, Nat. Photonics \textbf{2}, 496–500 (2008).
\bibitem{Alam} M. Alam, J.S. Aitchsion, and M. Mojahedi, Appl. Opt. \textbf{50}, 2294 (2011).
\bibitem{Yang} Yi Yang \textit{et al}, Nano Lett. \textbf{17} 3238 (2017).
\bibitem{leperchec09} J. Le Perchec, Y. Desieres, and R. Espiau de Lamaestre, Appl. Phys. Lett. \textbf{94}, 181104 (2009).
\bibitem{Cui} Y. Cui \textit{et al}, Laser Photonics Rev. \textbf{8}, 495-520 (2014).
\bibitem{LePerchec11} J. Le Perchec \textit{et al}, Opt. Express \textbf{19}, 15720-15731 (2011).
\bibitem{Li} T. Li, V. Nagal, D.H. Gracias, and J.B. Khurgin, Opt. Lett. \textbf{43}, 4465 (2018).
\bibitem{Devilez} A. Devilez \textit{et al},  Phys. Rev. B \textbf{92}, 241412 (R) (2015).
\bibitem{LePerchec08} J. Le Perchec \textit{et al}, Phys. Rev. Lett. \textbf{100}, 066408 (2008).
\bibitem{Polyakov} A. Polyakov \textit{et al}, Scientific Reports \textbf{2}, 933 (2012).
\bibitem{Liu10} Y. Liu \textit{et al}, Opt. Express \textbf{18}, 25029 (2010).
\bibitem{Robinson} J. T. Robinson \textit{et al}, Phys. Rev. Lett. \textbf{95}, 143901 (2005).
\bibitem{Botten}L.C. Botten \textit{et al}, Optica Acta \textbf{28}, 413-428 (1981).
\bibitem{Trivedi} K. Trivedi et al, J. Vac. Sci. Technol. B \textbf{27}, 3145 (2009).
\bibitem{Tournier} A. Tournier \textit{et al}, Proc. Int. Image Sensor Workshop, pp. 12-15 (2011).
\bibitem{Leperchec10} J. Le Perchec, Europhys. Lett. \textbf{92}, 67006 (2010).
\bibitem{Quemerais} P. Qu\'emerais \textit{et al}, J. Appl. Phys. \textbf{97}, 053507 (2005).


\end{thebibliography}
\end{document}